\documentclass[prl,twocolumn,showpacs,a4paper]{revtex4}
\usepackage{graphicx}
\usepackage{amsmath}

\begin{document}

\newcommand\ung{\ensuremath{\sim}}
\newcommand {\da} {\ensuremath{d^\dagger}}
\newcommand {\dnn} {\ensuremath{d^{\phantom{\dagger}}}}
\newcommand {\ca} {\ensuremath{c^\dagger}}
\newcommand {\cnn} {\ensuremath{c^{\phantom{\dagger}}}}
\newcommand{\nbr} {\ensuremath{\langle ij \rangle}}

\title{Observation of Elastic Doublon Decay in the Fermi-Hubbard Model}

\author{Niels Strohmaier}
\affiliation{Institute for Quantum Electronics, ETH Zurich, 8093 Zurich,
Switzerland}

\author{Daniel Greif}
\affiliation{Institute for Quantum Electronics, ETH Zurich, 8093 Zurich,
Switzerland}

\author{Robert J\"ordens}
\affiliation{Institute for Quantum Electronics, ETH Zurich, 8093 Zurich,
Switzerland}

\author{Leticia Tarruell}
\affiliation{Institute for Quantum Electronics, ETH Zurich, 8093 Zurich,
Switzerland}

\author{Henning Moritz}
\email{moritz@phys.ethz.ch}
\affiliation{Institute for Quantum Electronics, ETH Zurich, 8093 Zurich,
Switzerland}

\author{Tilman Esslinger}
\affiliation{Institute for Quantum Electronics,
ETH Zurich, 8093 Zurich, Switzerland}

\author{Rajdeep Sensarma,$^{1,2}$ David Pekker,$^1$ Ehud Altman,$^3$ and Eugene
Demler$^1$}

\affiliation{$^1$Department of Physics, Harvard University, Cambridge,
Massachusetts 02138, USA}

\affiliation{$^2$Condensed Matter Theory Center, University of Maryland, College Park, Maryland 20742, USA}

\affiliation{$^3$Department of Condensed Matter Physics, Weizmann Institute,
Rehovot, 76100, Israel}

\date{\today}

\begin{abstract}
  We investigate the decay of highly excited states of ultracold
  fermions in a three-dimensional optical lattice. Starting from a
  repulsive Fermi-Hubbard system near half filling, we generate
  additional doubly occupied sites (doublons) by lattice modulation.
  The subsequent relaxation back to thermal equilibrium is monitored
  over time. The measured absolute doublon lifetime covers two orders of magnitude.
  In units of the tunneling time $h/J$ it is found to depend
  exponentially on the ratio of on-site interaction energy $U$ to
  kinetic energy $J$. We argue that the dominant mechanism for the
  relaxation is a simultaneous many-body process involving several
  single fermions as scattering partners.
  A many-body calculation is carried out using diagrammatic methods, yielding
  fair agreement with the data.
  \end{abstract}


\maketitle


Understanding the far-from-equilibrium dynamics of strongly
correlated systems is a highly challenging task. Even the
identification of the basic processes involved and the associated
time scales is nontrivial when the system cannot be described by
weakly interacting excitations or quasiparticles. In these systems,
dynamics may couple states with widely different energies making the
description in terms of a restricted set of low energy states
impossible. While progress has been achieved for one-dimensional
systems (\cite{DMRG,Kinoshita2006}, and references therein), these
results can typically not be extended to higher dimensions.

The main difficulty in analyzing non-equilibrium dynamics in the
setting of condensed matter experiments is the strong coupling to
the environment, which introduces extrinsic relaxation mechanisms
and makes it challenging to prepare far-from-equilibrium initial
states in a controlled way. By contrast, the nearly perfect
isolation of many-body systems realized with ultracold atoms makes
them a perfect candidate for studying the intrinsic dynamics of
strongly correlated systems. In the setting of ultracold atoms it is
possible to prepare a well-controlled initial state, evolve it under
the action of a precisely defined microscopic Hamiltonian, and
monitor the effects of the characteristic relaxation process
\cite{Bloch2008}.

In this Letter, we take advantage of the recent realization of the
repulsive Fermi-Hubbard model with ultracold atom
systems~\cite{Kohl2005, Jordens2008, Schneider2008} to investigate
the relaxation of artificially created highly excited states. This
problem appears in diverse contexts like multiphonon decay of
excitons in semiconductors~\cite{Perebeinos2008}, pump-probe
experiments~\cite{PumpProbe} and dynamics of resonances in nuclear
matter~\cite{Brown1975}. Due to the negligible coupling to an
external environment, we are able to carry out a direct comparison
of experiment and theory. The interpretation of these results shows
the importance of high-order scattering processes in bridging the
energy gap between low- and high-energy excitations and how they can
lead to exponentially slow thermalization.

In the experiment, we study the time evolution of doubly occupied
lattice sites (doublons) in the repulsive Fermi-Hubbard model. This
model describes fermionic particles hopping on a lattice with
tunneling $J$ and on-site repulsion $U$ and is realized by a
two-component Fermi gas in an optical lattice.
In the context of a dilute Bose-Hubbard system isolated repulsively
bound pairs have been experimentally identified and
studied~\cite{Winkler2006}.

\begin{figure}[htbp]
  \includegraphics[width=0.9\columnwidth,clip=true]{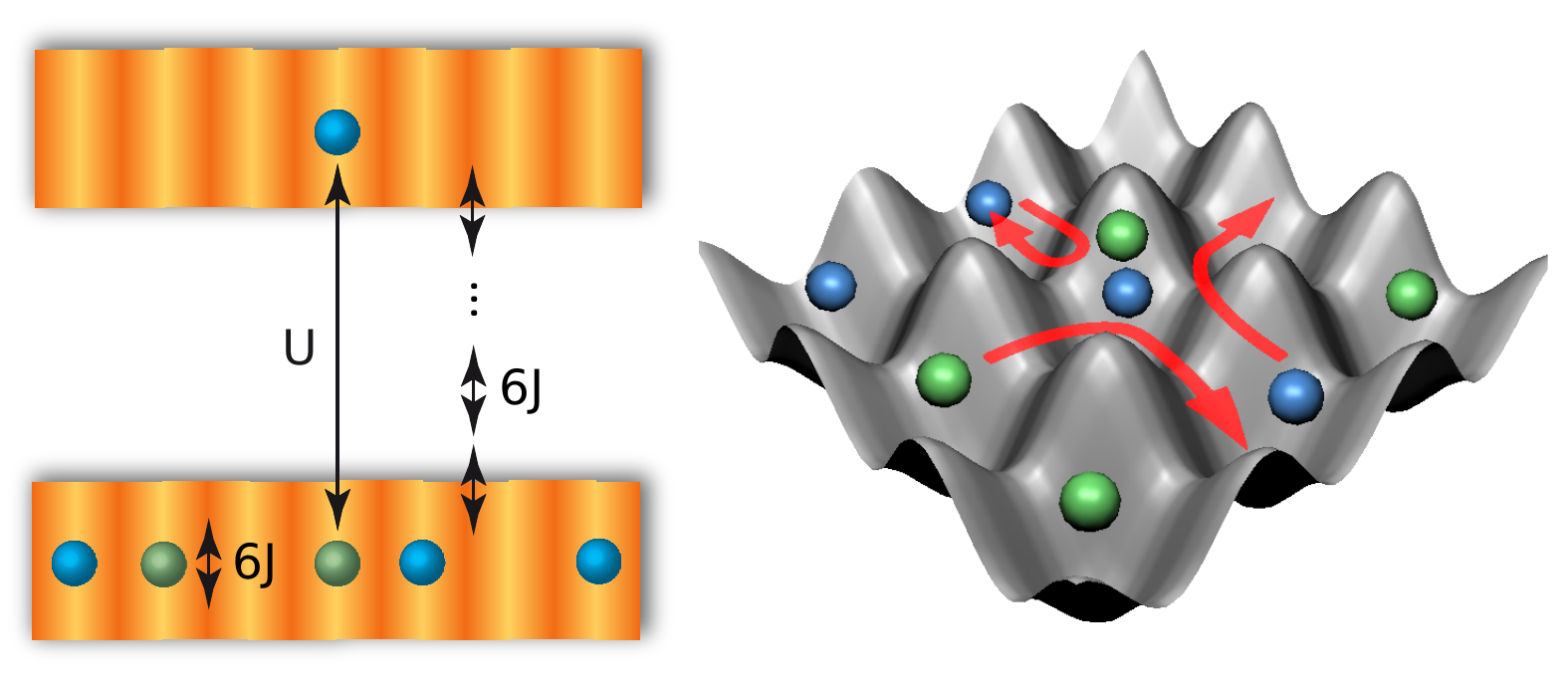}
   \caption{Stability of highly excited states in the single-band Hubbard model.
   Doubly occupied lattice sites are protected against decay by the on-site
   interaction energy $U$. The average kinetic energy of a single particle in a
   periodic potential is half the bandwidth $6J$. Thus the relaxation of
   excitations requires several scattering partners to maintain energy conservation.}
  \label{fig1:scetch}
\end{figure}

We report on the observation of elastic decay of artificially
created doublons~\cite{Jordens2008, Huber2008, Sensarma2009} into
single particles. The resulting lifetime is found to increase
exponentially with the ratio $U/6J$ (i.e. the lifetime becomes
longer as the interactions become stronger).
We argue that a doublon, having an excess energy $U$, decays in a
scattering process involving several single fermions,
cf. Fig.~\ref{fig1:scetch}. Since each of these scattering partners
can only absorb an average energy of $6J$, the number of virtual
states involved in the simultaneous many-body process is $U/6J$.
Hence the decay is exponentially suppressed for increasing $U/6J$.
We find fair agreement with diagrammatic calculations where the
strongly correlated nature of the underlying state is crucial in
obtaining the correct value of the scaling exponent.


The experimental sequence used to produce quantum degenerate Fermi gases has
been described in detail in Ref. \cite{Jordens2008}. In brief, we
prepare $(50\pm 5)\times10^3$ $^{40}$K atoms at temperatures below $15\%$ of the
Fermi temperature $T_{\mathrm{F}}$ in a balanced mixture of two magnetic
sublevels of the $F=9/2$ manifold. The confinement is given by a dipole trap
with trapping frequencies
$\omega_{x,y,z}=2\pi\times(35, 23, 120)$ Hz. Using Feshbach resonances in either
a $\left(m_{\mathrm{F}}=-9/2,-7/2\right)$ or $\left(m_{\mathrm{F}} =
-9/2,-5/2\right)$ mixture \cite{Regal2003, fbrnote}, the interaction strength
is tuned in the range $98\, a_0-131\, a_0$ or $374\, a_0-672\, a_0$
respectively, where $a_0$ is the Bohr radius. After adjusting the scattering
length to the desired value, we add a three-dimensional cubic optical lattice.
The lattice depth is increased in $200$\,ms to final values between
$6.5\,E_{\mathrm{R}}$ and $12.5\,E_{\mathrm{R}}$ in units of the recoil energy
$E_{\mathrm{R}}={h^2}/{2m\lambda^2}$. Here $\lambda=1064\,\text{nm}$ is the wavelength of the lattice beams.
The lattice beams have Gaussian profiles with $1/e^2$ radii of $w_{x,y,z}=
(160, 180, 160)\,\mathrm{\mu m}$. For a
given scattering length and lattice depth, $J$ and $U$ are inferred from Wannier
functions \cite{Jaksch1998}. Their statistical and systematic errors
are dominated by the lattice calibration and the accuracies in width and
position of the two Feshbach resonances \cite{Regal2003, fbrnote}. Depending on
$U$ and $J$ the accessible final regimes of the system range from metallic to
Mott insulating phases with a double occupancy below $15\%$.

The preparation of the system is followed by a sinusoidal modulation
of the lattice depth with an amplitude of $10\%$ and frequency close
to $U/h$. This causes an increase of the double occupancy to values up to $35\%$ \cite{Kollath2006,Hassler2008}.

After the modulation the system is in a non-equilibrium state, which we let
evolve freely at the initial lattice depth and interaction strength for up
to $4$\,s. This is followed by a sudden increase of the lattice depth to
$30\,E_{\mathrm{R}}$, which prevents further tunneling. We then measure the
amount of atoms residing on singly (doubly) occupied sites $N_{\mathrm{s}}$
($N_{\mathrm{d}}$) by encoding the double occupancy into a previously
unpopulated spin state using RF spectroscopy \cite{Jordens2008}. Combining
Stern-Gerlach separation and absorption imaging we obtain the single occupancy
$n_{\mathrm{s}}=N_{\mathrm{s}}/N_{\mathrm{tot}}$, double occupancy
$n_{\mathrm{d}}=N_{\mathrm{d}}/N_{\mathrm{tot}}$ and total atom number
$N_{\mathrm{tot}}=N_{\mathrm{s}}+N_{\mathrm{d}}$.

\begin{figure}[htb]
  \includegraphics[width=0.95\columnwidth,clip=true]{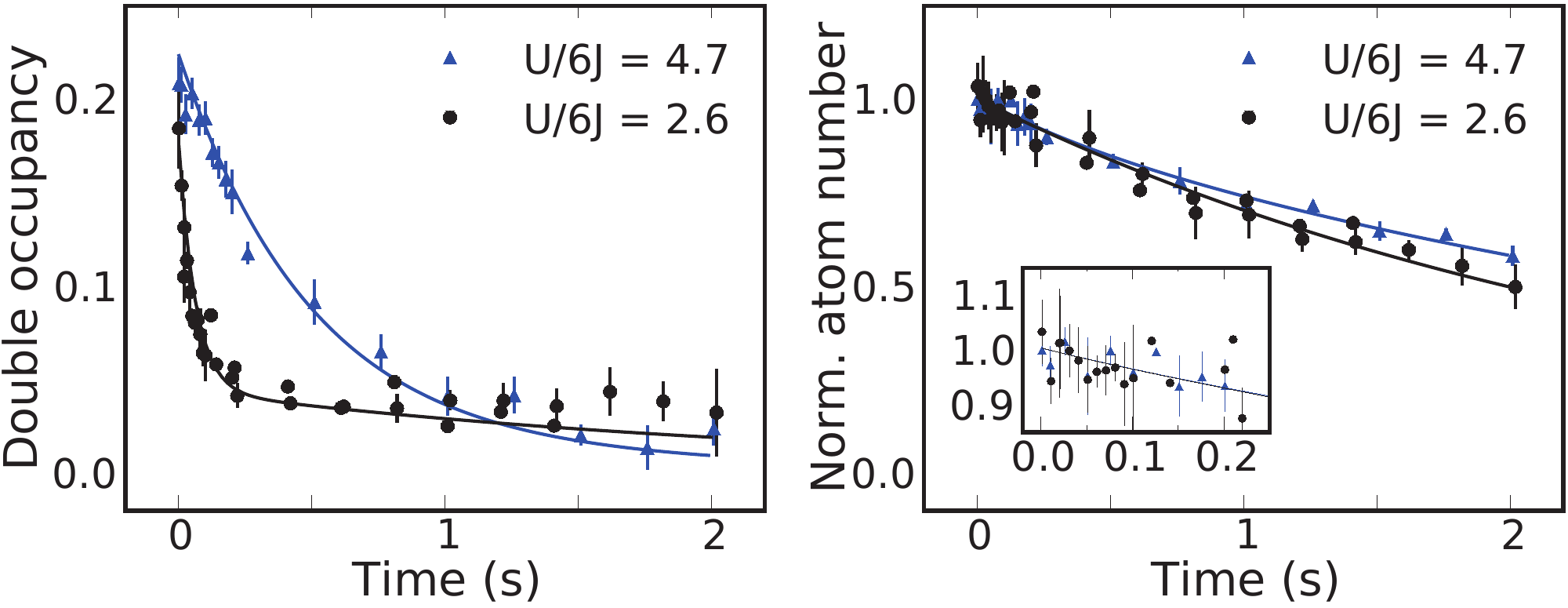}
   \caption{Comparison of the time evolution of the double occupancy and total
   atom number for different ratios $U/6J$. The data was recorded using the
   $(-9/2,-5/2)$ spin mixture, with $U/h=3.9$ kHz ($3.1$ kHz) and $J/h=140$ Hz
   ($200$ Hz) for the triangular (round) data points. The lines show fits of the
   integrated population equations of Eq.~\ref{eqn:popdecay}. The
   total atom numbers are scaled to the initial values. The inset shows a
   magnification for short times. Error bars denote the statistical error
   of at least four identical measurements.}
  \label{fig2:sampledecay}
\end{figure}

We record the time evolution of the total atom number and the single and double
occupancy, see Fig.~\ref{fig2:sampledecay}. The double occupancy is found to
decay exponentially, while additional losses are also observed on longer
timescales, which lead to a reduction of the total atom number. To extract the
doublon lifetime we model these decays by a set of coupled rate equations:
\begin{eqnarray}
\label{eqn:popdecay}
\Delta\dot{N}_{\mathrm{d}}&=&-\left(\frac{1}{\tau_{\mathrm{D}}}+\frac{1}
{\tau_{\mathrm{in}}}+\frac{1}{\tau_{\mathrm{loss}}}\right)\Delta
N_{\mathrm{d}}\nonumber\\
\dot{N}_{\mathrm{d},0}&=&-\left(\frac{1}{\tau_{\mathrm{in}}}+\frac{1}
{\tau_{\mathrm{loss}}}\right)N_{\mathrm{d},0}\\
\dot{N}_{\mathrm{s}}&=&\frac{1}{\tau_{\mathrm{D}}}\Delta N_{\mathrm{d}} -
\frac{1}{\tau_{\mathrm{loss}}}N_{\mathrm{s}}\nonumber
\end{eqnarray}
Here $\Delta N_{\mathrm{d}}$ is the additional amount of double
occupancy created by the lattice modulation as compared to the
equilibrium population $N_{\mathrm{d},0}$.
The three time constants correspond to three independent local decay
processes differing in the type of site they affect:
 the lifetime of doublons $\tau_{\mathrm{D}}$ describes a population flow from doubly occupied to singly occupied lattice sites visible as a fast decay (rise) of double (single) occupancy within $0.01-1$\,s. The other two times denote loss time constants, which lead to a reduction of the total atom number:
$\tau_{\mathrm{loss}}$ corresponds to losses affecting both site
types in the same manner, which is only observed in the total atom
number. Additional inelastic losses on doubly occupied sites are
summarized by $\tau_{\mathrm{in}}$, visible as a simultaneous decay
of both the total atom number and double occupancy. Changes of the
decay times during the decay and higher order terms in the rate
equations are excluded.

We simultaneously fit the time-dependent populations obtained from
Eq.~\ref{eqn:popdecay} to this dataset and to a corresponding
reference dataset without lattice modulation. Since the modulation
does not change the losses, this procedure removes the influence of
$\tau_\text{in}$ and $\tau_\text{loss}$, allowing for a reliable
determination of the doublon lifetime $\tau_\text{D}$. The model and
the observation are found to agree very well within experimental
uncertainties, as shown in Fig.~\ref{fig2:sampledecay}.

\begin{figure}[b]
  \includegraphics[width=0.9\columnwidth,clip=true]{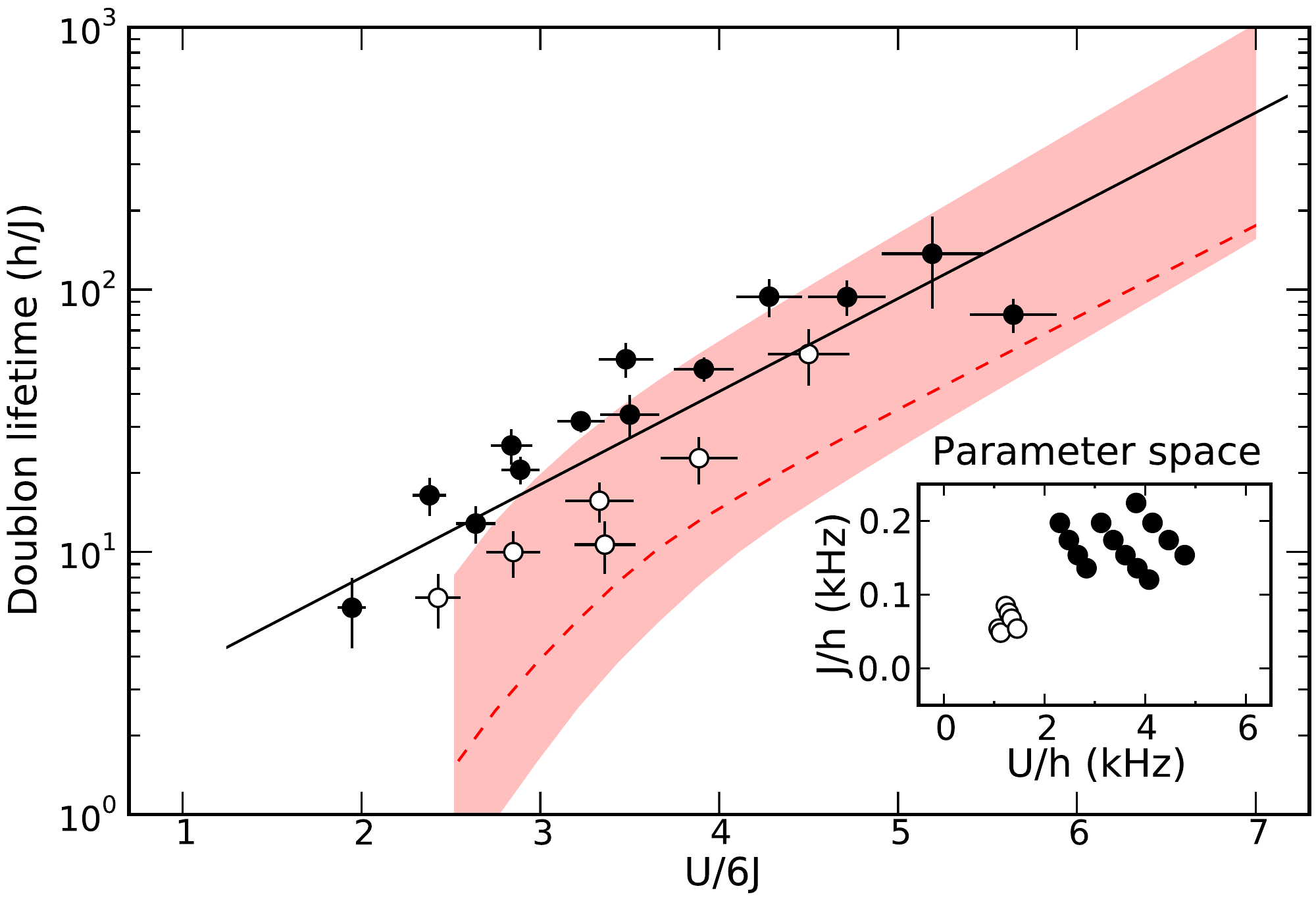}
   \caption{Semilog plot of doublon lifetime $\tau_\mathrm{D}$ vs. $U/6J$. The lifetime is extracted from datasets
   as shown in Fig.~\ref{fig2:sampledecay}. Solid and hollow circles
   denote the $\left(-9/2,-5/2\right)$ and $\left(-9/2,-7/2\right)$ spin
   mixture respectively, while the dashed line shows the theoretical result at half filling.
   The solid line is a fit of Eq.~\ref{scalingeqn} to the experimental data,
   yielding $\alpha=0.82\pm0.08$, whereas for the theory curve the asymptotic
   slope at large $U/6J$ is $\alpha_\mathrm{T}=0.80$. The shaded corridor was obtained
   by varying the filling factor in the calculation by $\pm0.3$ (which has only a weak
   effect on the slope). The inset shows the parameters used to realize the different values of
   $U/6J$. Error bars denote the confidence intervals of the lifetime fits and
   the statistical errors in $U/6J$. The systematic errors in
   $U/6J$ and $\tau_\text{D}/(h/J)$ are estimated to be $30\%$ and $25\%$,
   respectively.}
  \label{fig3:scalingdata}
\end{figure}

We measure this doublon lifetime for various tunneling and
interaction strengths, covering a parameter range where $J$ and $U$
each differ by at least a factor of four (inset Fig.~\ref{fig3:scalingdata}).
The lifetime in units of the tunneling time is plotted
logarithmically versus the ratio $U/6J$ in
Fig.~\ref{fig3:scalingdata}. The data is well described by an
exponential function:
\begin{equation}
\label{scalingeqn}
\frac{\tau_\mathrm{D}}{h/J} =C\, \exp\left(\alpha\frac{U}{6J}\right).
\end{equation}
The scaling exponent $\alpha$ is found to be $\alpha=0.82\pm0.08$
with $C=1.6\pm0.9$. We find fair agreement with our calculations of the
doublon lifetime. The systematic deviation
of the data for the two spin mixtures \cite{SeparateFits} seems
to indicate that the data show physics beyond Eq. (2).


In the following we argue that this exponential scaling of the doublon
lifetime originates from a high order scattering process involving
several single atoms as scattering partners. In the preparation of the
non-equilibrium state by lattice modulation,
we create holes as well as doublons in the bulk and thus drive the
system into a compressible state. An isolated doublon has an energy
$U$, which it must transfer to other excitations in order to decay. In
the compressible state the most relevant excitations are metallic with a typical
energy scale of $6J$. Thus a doublon must scatter with several fermions. The
number of scattering partners is on the order of $n=U/6J$. The matrix element
$M$ for the decay rate $\Gamma$ may be estimated via perturbation theory $M \ung\:\,\frac{J}{6J}\times\frac{J}{2\times6J}\times\dots\times\frac{J} {n\times6J}$ and $\Gamma/J \propto M^2$.
Using Stirling's formula, we then find the same scaling behavior as
in Eq.~\ref{scalingeqn}. Here $\alpha$ is a parameter on the order
of unity and depends at most logarithmically on $U/6J$.

For the quantitative analysis a few assumptions are made: we consider
the decay of a single doublon in the background of a homogeneous
compressible system. This is justified since most of the doublons are
created in the central region of the trap, where the filling is
highest, and decay at most within a few sites of where they are
produced (the estimated travel distance for a random walk during the decay
process is not more than $\sqrt{\tau_{\mathrm{D}} J/h} \lesssim 10$ sites,
which is less than the cloud radius). We neglect spin excitations and
collisions between doublons, as typical energy transfers in these processes
are on the order of $J^2/U$, which leads to a subdominant exponential scaling
in $U^2/J^2$. Further, the population of higher bands can be excluded, since
$U$ is always smaller than half the band gap. We also note that confinement
assisted decay of doublons after quantum tunneling to the edge of the cloud
is unlikely, as the accessible confinement energy is smaller than $U$ and the
tunneling rate is very small.

The complete Hamiltonian of the system may be written as $H=
H_\text{pf} +H_\text{d}+H_\text{fd}$, where $H_\text{pf}$ describes
the background fermions, $H_\text{d}$ is the on-site energy of
doublons and $H_\text{fd}$ is the interaction of the doublon with the
background fermions.

The strong Hubbard repulsion between the fermions leads to the concept of
projection, where two fermions are forbidden from occupying the same site.
In this case, the fermions can only hop in the presence of a hole on a
neighboring site and are governed by the Hamiltonian
\begin{align}
H_\text{pf}=-J \sum_{\langle i j\rangle, \sigma}
(1-n_{i,\bar{\sigma}}) c^\dagger_{i,\sigma} c_{j,\sigma} (1-n_{j,\bar{\sigma}}),
\end{align}
where $c^\dagger_{i,\sigma}$ ($c_{i,\sigma}$) is the fermion creation
(annihilation) operator and $n_{i,\sigma}$ is the number operator for
fermions with spin $\sigma$ ($\bar{\sigma}$ denotes spin opposite to
$\sigma$). Expanding out this Hamiltonian we obtain
$H_\text{pf}=H_\text{f}+H_\text{p}$, with
\begin{align}
H_\text{f}&=-J  \sum_{\langle i j\rangle, \sigma} c^\dagger_{i,\sigma}
c_{j,\sigma} - \mu \sum_{i,\sigma} c^\dagger_{i,\sigma}
c_{i,\sigma} ,\\
H_\text{p}&=J \sum_{\langle i j\rangle, \sigma}
\left(
n_{i,\bar{\sigma}} c^\dagger_{i,\sigma} c_{j,\sigma}+
c^\dagger_{i,\sigma} c_{j,\sigma} n_{j,\bar{\sigma}}
\right),
\end{align}
where $H_\text{f}$ describes the free Fermi sea and $H_\text{p}$
describes the interaction induced by the projection and can be thought
of as a process in which a fermion scatters off the Fermi sea and
creates a particle-hole pair. We assume that the system is close to half filling (chemical potential $\mu=0$),
but we checked that the result is not very sensitive to the precise value of the filling as shown by the shaded region in Fig.~\ref{fig3:scalingdata}.
We neglect the term
$n_{i,\bar{\sigma}}c^\dagger_{i,\sigma}c_{j,\sigma}n_{j,\bar{\sigma}}$
in $H_\text{p}$ as we have checked that it leads to negligibly small corrections to the
doublon decay rate~\cite{usLong}.

We now consider the propagation and decay of a doublon in the
background state of the projected Fermi sea. The on-site energy of the
doublon is $H_\text{d}=U\sum_i d^\dagger_i d_i$, where
$d^\dagger$ is a doublon creation operator. The doublon-fermion
interaction $H_\text{fd}$ is given by
\begin{align}
H_\text{fd}&\!=\!\! J \sum_{\nbr} (\da_i \dnn_i+\da_j \dnn_j+\da_j \dnn_i) \ca_{i\sigma}\cnn_{j\sigma} \\\nonumber
&\quad\quad\quad\quad\quad + d_i (\ca_{i\uparrow}\ca_{j\downarrow}\!-\!\ca_{i\downarrow}\ca_{j\uparrow})+ \text{h.c.},
\end{align}
where the terms describe: projecting out configurations with a
doublon and a fermion on the same site (first and second term), hopping of
doublons with back-flow of fermions (third term) and interconversion between
a pair of single fermions and a doublon (last term).

We now see that there are two different processes by which the doublon
can lose energy. It can create a large number of particle-hole pairs
(through the first term in $H_\text{fd}$), each with an energy of on
the order $6J$, or it can create a high energy particle-hole pair
(through $H_\text{fd}$), which is itself unstable and decays into a
shower of particle-hole pairs (through the action of
$H_\text{p}$). The last process is the result of strong interaction between the fermions and must be taken into account in order to obtain an accurate
estimate of the doublon lifetime.
\begin{figure}[t]
  \includegraphics[width=6cm]{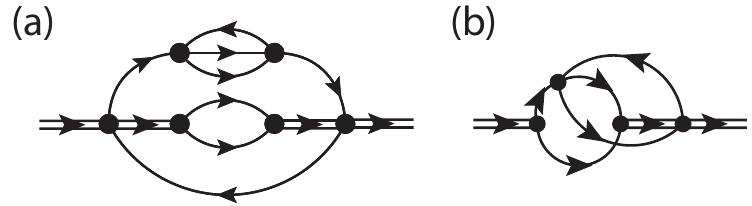}
  \caption{The double lines represent doublon propagators, and the
    single lines fermion propagators. (a)~Typical doublon propagator
    diagram showing the creation of particle-hole pairs by both the
    doublon and the projected fermions as well as annihilation of the
    doublon into a pair of single fermions. (b)~Typical example for a neglected
    diagram type.}
  \label{fig4:diagram}
\end{figure}

Our strategy for determining the doublon lifetime is to compute the
doublon self-energy $\Sigma\left(\omega\right)$ diagrammatically
\cite{usLong} and obtain the decay rate from $\text{Im}\,
\Sigma\left(U\right)$, the imaginary part of the self-energy at
$\omega = U$.  We proceed by first obtaining the Green function for the
projected Fermi sea ($H_{\rm{f}}+H_{\rm{p}}$) using a diagrammatic perturbation
theory. Next, we use this Green function in a resummation procedure to
obtain $\Sigma\left(\omega\right)$. These steps can be treated as
independent when the doublon density is small, as the presence of the
doublons does not change the background fermion Green's functions.
Throughout, we follow the principle of maximizing the number of
particle-hole pairs (see Fig.~\ref{fig4:diagram}a) at each order of
perturbation theory. We do miss the class of diagrams in which
interactions between fermions cannot be described by a fermion
self-energy (see Fig.~\ref{fig4:diagram}b). We carry out our calculations
in the zero temperature formalism. However, since we are looking at
high energy processes ($\omega \ung\:U$), finite temperatures will not
have a large effect on the results as long as $T\leq U$ \cite{usLong}.

Our theoretical analysis was constructed to capture the scaling
parameter of the doublon lifetime at large $U/6J$, as it relies on
generating a large number of particle-hole pairs. In this regime the
theoretically computed value of the scaling exponent is
$\alpha_\text{T}=0.80$ close to half filling, which agrees well with
the experimentally obtained value $\alpha=0.82\pm0.08$. For small
$U/6J$ the theory breaks down, leading to disagreement between
experiment and theory in this regime (see
Fig.~\ref{fig3:scalingdata}). Although the theory is not designed to
predict the pre-exponential factor $C$, we find reasonable agreement
between theory and experiment. For large $U/6J$ losses are expected
to mask the observation of very long lifetimes.


In conclusion, we have investigated the non-equilibrium dynamics of
fermions in an optical lattice and shown that the lifetime of
doublons scales exponentially with the ratio of interaction energy
to kinetic energy. We argue that the dominant decay mechanism of
doublons is a high order scattering process involving several single
particles, and we obtain fair agreement with the experiments based
on a perturbation theory calculation. The results have implications
also for the simulation of strongly correlated lattice models with ultracold
atoms as they pose adiabaticity constraints on the sweep rates for the system
parameters. On a qualitative level, the results should also be
applicable to bosonic atoms and might help to explain the long
equilibration times recently observed in \cite{Chin2009}.

We thank A. Georges and A. Rosch for insightful discussions, and
SNF, NAME-QUAM (EU) and SCALA (EU) for funding. R. S., D. P., and E.
D. acknowledge the support of NSF, DARPA, MURI and CUA. E. A.
acknowledges support from BSF (ED, EA) and ISF.

\end{document}